\documentclass[aps,prd,reprint,showpacs]{revtex4-1}

\usepackage{amsmath}
\usepackage{amssymb}

\begin{document}

\newcommand{\e}{{\rm e}}
\newcommand{\pa}{\partial}
\newcommand{\parl}{\!\buildrel \leftrightarrow\over\pa\!\!}
\newcommand{\rmd}{{\rm d}}
\newcommand{\ve}{\varepsilon}
\newcommand{\ts}{\textstyle}
\newcommand{\x}{\;\!}
\newcommand{\beq}{\begin{eqnarray}}
\newcommand{\eeq}{\end{eqnarray}}
\newcommand{\ba}{\begin{array}}
\newcommand{\ea}{\end{array}}
\newcommand{\nn}{\nonumber}
\newcommand{\no}{\nonumber\\}
\title{The Unruh Effect Revisited}
\author{Pietro Longhi}
\email{longhi@physics.rutgers.edu}
\affiliation{Department of Physics and Astronomy\\
Rutgers, The State University of New Jersey\\
136 Frelinghuysen Road,
Piscataway, NJ 08854-8019 USA}
\author{Roberto Soldati}
\email{roberto.soldati@bo.infn.it}
\affiliation{Dipartimento di Fisica, Universit\`a
di Bologna\\ 
Istituto Nazionale di Fisica Nucleare, Sezione di Bologna\\
Via Irnerio 46, 40126 - Bologna (Italy)}
\date{\today}
\begin{abstract}
A new and exact derivation of the Bogoliubov coefficients is obtained
for the simplest case of a spinless, neutral, massive field in
a uniformly accelerated frame with  a constant acceleration.
The method can be suitably generalized in a straightforward manner
to any field with spin and charges.
\end{abstract}
\maketitle
The so called Unruh effect \cite{Unruh} has been discovered long ago and since
its very appearance a really large number of investigations has been developed to the aim 
of applying and generalizing its
deep and seminal content to a wide range of physical contexts, such as e.g. black holes, dark matter and
dark energy, cosmological horizons, acoustic black holes \textit{et cetera}. 
A very nice and recent review paper 
\cite{crispino} on this
subject actually provides a pretty good taste of the impressive growth of interest
on this subject during the last few years, including the possible occurrence \cite{balbo} of its
experimental observation.
From the field theoretical point of view, the Unruh effect is encoded in the
Bogoliubov coefficients that relate the complete and orthonormal sets of the wave functions,
as experienced by two observers moving respectively in an inertial reference frame and in
a noninertial accelerated one. Curiously enough, as reported in \cite{crispino},
the evaluation of the Bogoliubov coefficients in the case of a constant relative acceleration
has never been done directly from the very definition but arguing indirectly, sometimes in an
admittedly tricky manner \cite{takagi}.

It is the aim of the present note to fill that lack, i.e., we will perform  a 
straightforward exact calculation of the Bogoliubov coefficient from field theoretic first principles.
To start with, we briefly review the canonical quantization of a spinless and chargeless field 
in a Rindler's coordinate system, i.e., as experienced by some uniformly accelerated
noninertial observer. In so doing we shall establish our notations and conventions.
Consider
the four dimensional Minkowski spacetime with the line element \cite{BD}
\[
\mathrm d s^2=\eta_{\,\alpha\beta}\,\mathrm d X^{\alpha}\mathrm d X^{\beta}
=g_{\,\mu\nu}(\mathrm x)\,\mathrm d \mathrm x^{\,\mu}\mathrm d \mathrm x^{\,\nu}
\]
where the constant metric tensor $\eta_{\,\alpha\beta}={\rm diag}\,(+,-,-,-)$
is relative to an inertial coordinate system in Minkowski spacetime and
labelled by the so called anholomic indices denoted with the greek letters 
from the first part of the alphabeth, while the
Einstein convention on the sum over repeated indices is understood.
We employ natural units $\hbar=c=1$ unless explicitly
stated. If we set
\begin{equation}
X^{\alpha}=\left(\,\tau,X,y,z\,\right)\qquad\quad
\mathrm x^{\,\mu}=(t,x,y,z)
\end{equation} 
then we shall denote the following 
spacelike region of the Minkowski spacetime
\begin{equation}
{\mathfrak W}_R = \{X^\mu\in{\mathbb R}^4\,|\,X\ge0\,,\,\tau^2\le X^2\}
\end{equation} 
as the right Rindler wedge.
Here we introduce the so called Rindler curvilinear coordinate system,
which describes an accelerated observer: namely,
\begin{align}
  &\tau=x\sinh({\rm a} t)\qquad\quad
(\,{\rm a}>0\,,\ \xi\in\mathbb R\,)\label{RRW0}\\
  &X=x\cosh({\rm a} t)
\qquad\qquad\quad\
(\,x\ge0\,)\label{RRW1}
\end{align}
where ${\rm a}$ is the constant acceleration.
The above coordinate transformations
can be readily inverted, viz.,
\begin{equation}
t={\rm a}^{-1}\,{\rm Arth}\left(\,\tau/X\,\right)\\
x=\sqrt{X^2 - \tau^2}\qquad\ \  (\,x\ge0\,)
\end{equation} 
in such a manner that we can also write
\begin{equation}
\mathrm{d} s^2=g_{\,\mu\nu}(x)\,\mathrm{d} \mathrm{x}^{\,\mu}\mathrm{d}\mathrm{x}^{\,\nu}
= x^2 {\rm a}^2\,\mathrm{d} t^2-\mathrm{d} x^2-\rmd y^2-\mathrm{d} z^2
\end{equation}
whence we obtain
\beq
g_{\,\mu\nu}(x)=
\left\lgroup
\ba{cccc}
{\rm a}^2x^2 & 0 & 0 & 0\\
0 & -\,1& 0 & 0\\
0 & 0 & -\,1 & 0\\
0 & 0 & 0 & -\,1 
\ea
\right\rgroup
\eeq
so that
\beq
g = {\rm det}\,g_{\,\mu\nu}(x)=[\,{\rm det}\,g^{\,\mu\nu}(x)\,]^{\,-1}=\,
-\,{\rm a}^2x^2
\eeq
Notice that, by changing both signes in the definitions
(\ref{RRW0}) and (\ref{RRW1}), we shall cover
the other spacelike region of the Minkowski spacetime,
i.e. the left Rindler wedge
\beq
{\mathfrak W}_L = \{X^\mu\in{\mathbb R}^4\,|\,X\le0\,,\,\tau^2\le X^2\}
\eeq 
Moreover we have the Christoffel symbols
\[
\Gamma_{\mu\nu}^{\lambda}(x)=
{\textstyle\frac12}g^{\,\kappa\lambda}(x)
\left\lbrace\pa_\mu g_{\,\nu\kappa}(x)+\pa_\nu g_{\,\mu\kappa}(x)
-\pa_\kappa g_{\,\mu\nu}(x)\right\rbrace
\]
so that there are only two nonvanishing components of the affine connection:
namely,
\begin{align}
&\Gamma_{01}^{\,0}(x)={\textstyle\frac12}\,g^{\,00}(x)\,\frac{\rm d}{\rmd x}\,g_{\,00}(x)
=\frac{1}{x}=\Gamma_{10}^{\,0}(x)\\
&\Gamma_{00}^{1}(x)={\frac12}\cdot\frac{\rm d}{\rmd x}\,g^{\,00}(x)={\rm a}^2 x
\end{align}
This entails the coordinate invariant Klein-Gordon second order differential operator
\begin{align}
& g^{\,\mu\nu}(x)\,D_\mu\pa_\nu + m^2 = g^{\,00}(x)\left(\pa_t^2-{\rm a}^2x\,\pa_x\right)-\Delta + m^2 \no
& \qquad=\frac{1}{x^2\rm a^2}\,\pa_t^2-\frac{1}{x}\cdot\pa_x-\pa_x^2-\pa_{y}^2-\pa_z^2 + m^2
\end{align}
where $\Delta$ is the Laplace operator, that leads to the invariant  Klein-Gordon equation
for a spinless field
\begin{equation}
 \left\lbrace g^{\,\mu\nu}(x)\,D_\mu\pa_\nu + m^2 \right\rbrace \phi(\mathrm x)=0
\label{KGE} 
\end{equation} 
The invariant scalar product between two solutions of the covariant Klein-Gordon field
equation is defined by \cite{BD}
\beq
(\,\phi_2\,,\,\phi_1\,)\equiv 
\oint_\Sigma\,\phi_2^\ast(\mathrm x)\,i\parl_\lambda\,\phi_1(\mathrm x)\,\rmd\Sigma^\lambda
\label{invinnpro}
\eeq
where $\Sigma$ is a 3-dimensional spacelike hypersurface while 
\begin{align}
&\rmd\Sigma^{\lambda}\equiv {\textstyle\frac16}\,
\ve^{\,\lambda\mu\nu\rho}\rmd x_\mu\rmd x_\nu\rmd x_\rho\,(-\,g)^{-1/2}\\
&= {\textstyle\frac16}\,
\ve^{\,\lambda\mu\nu\rho}\,g_{\,\mu\alpha}(x)\,g_{\,\nu\beta}(x)\,g_{\,\rho\sigma}(x)\,
\rmd x^\alpha\,\rmd x^\beta\,\rmd x^\sigma\,(-\,g)^{-1/2}\nn
\end{align}
is the invariant oriented hypersurface element with $\ve^{0123}=1\,.$
Thus, for the initial time 3-dimensional hypersurface 
in the right Rindler wedge
$\mathfrak W_R$ we get
\beq
\rmd\Sigma^{0}\,=\,-\;\frac{\theta(x)}{{\rm a}\,x}\;{\rmd x}\,\rmd^2{\bf x}_\perp
\qquad\rmd\Sigma^\imath=0\quad(\,\imath=1,2,3\,)
\eeq
where $\theta(x)$ is the Heaviside step distribution.
A complete and orthonormal set of stationary solutions
for the Klein-Gordon equation in the right Rindler wedge 
is provided by the generalization of the
so called Fulling modes \cite{fulling,fedotov}
\begin{align}
u_{E,\,{\bf k}}(t,x,{\bf x}_\perp)&=\frac{1}{2\pi^2\surd{\rm a}}\;
\sqrt{\sinh\left(\frac{\pi E}{{\rm a}}\right)}
\;K_{iE/\rm a}(\kappa x)\no
& \times \exp\{\,-iEt + i{\bf k}\cdot\,{\bf x}_\perp\}
\end{align}
where $K_\nu(x)$ is the modified Bessel function.
One can fix the normalization constant by the natural requirement
\beq
(\,u_{E^{\,\prime},\,{\bf k}^{\,\prime}}\,,\,u_{E,\,{\bf k}}\,)\,=\,
\delta(E-E^{\,\prime})\,\delta({\bf k}-{\bf k}^{\,\prime})
\eeq
Notice that the orthonormalization relation for the real modified Bessel functions
functions can be recast in the suggestive form
\begin{align}
\int_{0^+}^\infty\;\frac{\rmd x}{x}\,K_{i\nu^{\x\prime}}(\kappa x)\,
K_{i\nu}(\kappa x)&=\frac{\pi^2\delta(\nu-\nu^{\,\prime}\x)}{2\nu\sinh(-\x\pi\nu\x)}
\label{Bassett}
\end{align}
where $\nu\equiv{cE}/{\hbar\rm a}$.
Then we eventually get the complete and orthonormal set of the spinless and chargeless normal modes
in the right Rindler wedge $\mathfrak W_R$: namely,
\begin{align}
u_{\x\mbox{\boldmath $\nu$}}(t,x,{\bf x}_\perp)
&=\frac{1}{2\pi^2}
\;\sqrt{\rm a^{-1}\sinh\left(\pi\nu\right)}\,
K_{i\nu}(\kappa x)\no
&\times \exp\{\,-\x i\x{\rm a}\x\nu\x t + i\x{\bf k}\cdot\,{\bf x}_\perp\}\\
x>0\ \qquad & \qquad \kappa=\sqrt{{\bf k}^2+m^2}\qquad
\mbox{\boldmath $\nu$}\equiv (\nu,{\bf k})\label{ISNM}
\end{align}
which satisfy the invariant orthonormality relations
\beq
(\,u_{\x\mbox{\boldmath $\nu$}}\,,\,u_{\x{\mbox{\boldmath $\nu$}}^{\x\prime}}\,)=
{\rm a}^{-1}\;\delta(\x\mbox{\boldmath $\nu$} - {\mbox{\boldmath $\nu$}}^{\x\prime}\x)
\eeq
According to our definitions, the invariant scalar normal modes (\ref{ISNM}) 
have the standard canonical dimensions
$[\,u_{E,\,{\bf k}}\,]$ $= [\,u_{\x\mbox{\boldmath $\nu$}}\,]$ $= \sqrt{\rm cm}=$ eV$^{\,-1/2}$ 
in natural units.
It follows that the Fulling normal modes expansion of the
real scalar quantized field on the right Rindler wedge $\mathfrak W_R$
reads, in accordance with \cite{crispino,fulling,fedotov}
\begin{align}
\phi_R(\mathrm x)=\int_{-\infty}^\infty{\rmd\nu}\int\rmd^2{\bf k}\,
[\,a_{\x\mbox{\boldmath $\nu$}}\,u_{\x\mbox{\boldmath $\nu$}}(\mathrm x)
+a^{\x\dagger}_{\x\mbox{\boldmath $\nu$}}\,u^{\x\ast}_{\x\mbox{\boldmath $\nu$}}(\mathrm x)\,]
\end{align}
where
\begin{align}
&[\,a_{\x\mbox{\boldmath $\nu$}}\,,\,a_{\x{\mbox{\boldmath $\nu$}}^{\x\prime}}^{\,\dagger}\,]\,
=\,\mathrm a\,\delta(\nu-\nu^{\,\prime})\,\delta({\bf k}-{\bf k}^{\,\prime})\\
&[\,a_{\x\mbox{\boldmath $\nu$}}\,,\,a_{\x{\mbox{\boldmath $\nu$}}^{\x\prime}}\,]=0
=[\,a^{\dagger}_{\x\mbox{\boldmath $\nu$}}\,,\,a^{\dagger}_{\x{\mbox{\boldmath $\nu$}}^{\x\prime}}\,]
\end{align}
in such a way that the quantized real scalar field in the Rindler coordinates 
has the customary dimensions $[\,\phi\,]=$ eV in natural units, while
the operators $[\,a_{\x\mbox{\boldmath $\nu$}}\,]={\rm cm}^{3/2}$.
On the other side, the quantized real scalar Klein-Gordon field in the Minkowski spacetime
has the usual normal mode expansion
\beq
\phi_M(\tau,{\bf X})=\int\rmd{\bf K}\,[\,\bar a_{{\bf K}}\,\bar u_{\,{\bf K}}(\tau,{\bf X})
+ \bar a_{{\bf K}}^\dagger\,\bar u_{\,{\bf K}}^\ast(\tau,{\bf X})\,]
\eeq
where $K^{\alpha}\equiv\left(K_0,K,{\bf K}_\perp\right)$ and $K_0=\omega_{\,\bf K}=\sqrt{\displaystyle{\,{\bf K}^2 + m^2}}$ whereas
\begin{align}
&\bar u_{\,{\bf K}}(\tau,{\bf X})=[\,(2\pi)^3\,2\omega_{\,\bf K}\,]^{\,-1/2}\,
\exp\{-\,i\,K_\alpha\,X^\alpha\}\\
&(\,\bar u_{\,{\bf K}}\,,\,\bar u_{\,{\bf K}^{\,\prime}}\,)
=\delta({\bf K}-{\bf K}^{\,\prime})\quad\quad\quad
(\,\forall\,{\bf K},{\bf K}^{\,\prime}\in\mathbb R^3\,)
\end{align}
in which the creation and destruction operators satisfy the standard algebra
\begin{align}
& [\,\bar a_{\,{\bf K}}\,,\,\bar a_{\,{\bf P}}^\dagger\,]=\delta({\bf K}-{\bf P}) \\
& \nn [\,\bar a_{\,{\bf K}}\,,\,\bar a_{\,{\bf P}}\,]=0 \qquad[\,\bar a_{\,{\bf K}}^\dagger\,,\,\bar a_{\,{\bf P}}^\dagger\,]=0 \qquad 
\end{align}
that leads to the invariant canonical commutation relation
\begin{align}
&[\,\phi_M(\tau,{\bf X})\,,\,\phi_M(\tau^{\prime},{\bf X}^{\prime})\,]
=\frac{1}{i}\,D(\tau-\tau^{\prime},{\bf X}-{\bf X}^{\prime};m) \\
& \nn =\int\rmd{{\bf K}}\left\{\,\bar u_{\,{\bf K}}(\tau,{\bf X})\,
\bar u^{\ast}_{\,{\bf K}}(\tau^{\prime},{\bf X}^{\prime})
- \bar u^{\ast}_{\,{\bf K}}(\tau,{\bf X})\,
\bar u_{\,{\bf K}}(\tau^{\prime},{\bf X}^{\prime})\,\right\}
\end{align} 
where $D(\xi;m)$ is the 
invariant Pauli-Jordan distribution,
which is real, odd and enjoys the microcausality property
\begin{align}
& [\,\phi_M(\tau,{\bf X})\,,\,\phi_M(\tau^{\prime},{\bf X}^{\prime})\,]=
[\,\phi_R(t,\mathbf x)\,,\,\phi_R(t^{\,\prime},\mathbf x^{\,\prime})\,]=0 \no
& {\forall}\,X^{\x\mu}\,,\,X^{\x\prime\x\mu}\in\mathfrak W_R\,,\ (X-X^{\x\prime}\x)^2<0\\
& \nn \lim_{\tau^{\prime}\rightarrow\tau}\,\frac{i\partial}{\partial\tau}\,
[\,\phi_M(\tau,{\bf X})\,,\,\phi_M(\tau^{\prime},{\bf X}^{\prime})\,]=\\
&\lim_{\tau^{\prime}\rightarrow\tau}\,
\frac{\partial}{\partial\tau}\,D(\tau-\tau^{\prime},{\bf X}-{\bf X}^{\prime};m)=
\delta({\bf X} - {\bf X}^{\prime})
\end{align}
Of course it is apparent that, as far as a quantized scalar field is involved,
the canonical commutation relations in the right Rindler wedge
are frame independent, even for accelerated noninertial observers
since, by definition, 
\begin{equation}
 \phi_M[\,\tau(t,x), X(t,x),y,z\,]=\phi_R(\mathrm x)\qquad\quad\forall\,\mathrm x\in\mathfrak W_R
\end{equation}
Things are different and far more tricky for quantized fields with nonvanishing spin.

\bigskip
We now derive the exact structure of the Bogoliubov coefficients for a neutral spinless field,
leading to the well known Unruh's effect. To this purpose,
we have to deal with two complete orthonormal sets of solutions of the Klein-Gordon equation
(\ref{KGE}) in the right Rindler wedge and in the Rindler curvilinear coordinate system: namely,
\begin{align}
u_{\x\mbox{\boldmath $\nu$}}(\mathrm x)\,&=\,\frac{1}{2\pi^2}
\;\sqrt{\rm a^{-1}\sinh\left(\pi\nu\right)}\,K_{i\nu}(\kappa x)\\
&\times\exp\{\,-\x i\x{\rm a}\x\nu\x t + i\x{\bf k}\cdot\,{\bf x}_\bot\}\no
\bar u_{\,{\bf K}}(\mathrm x)&=[\,(2\pi)^3\,2\omega_{\,\bf K}\,]^{-\frac12}\,
\exp\{i\,\Bbbk_t\,x + i\,\mathbf K_{\bot}\cdot\mathbf x_{\bot}\}\\
\Bbbk_t\equiv &K\cosh(\mathrm at) - \omega_{\,\bf K}\sinh(\mathrm at)
\qquad\quad [\,x>0\,,\,\nu\in\mathbb R\,] \nn
\end{align} 
Then a direct substitution into the invariant inner product (\ref{invinnpro}) yields
for $E=\mathrm a\nu$
\begin{align}
(\,\bar u_{\,{\bf K}}\,,\,u_{\x\mbox{\boldmath $\nu$}}\,) &=
\delta({\bf K}_\perp-{\bf k})\,\exp\{-\,iEt\}\,
\sqrt{\frac{2\sinh(\pi\nu)}{(2\pi\rm a)^3\,\omega_{\,\bf K}}}\no
&\times
\int_{0^+}^\infty\frac{\rmd x}{x}\,\exp\{-\,i\x\Bbbk_tx\}\,K_{iE/\mathrm a}(\kappa x)\,
\left( x \dot\Bbbk_t - E \right)
\end{align}
where $\dot\Bbbk_t=\mathrm aK\sinh(\mathrm at) - \mathrm a\omega_{\,\bf K}\cosh(\mathrm at)$.
However, since the invariant inner product is coordinate time independent by construction,
in accordance with the textbook notations \cite{BD} we can safely write
\begin{align}
(\,\bar u_{\,{\bf K}}\,,\,u_{\,\mbox{\boldmath $\nu$}}\,)&=
-\,\delta(\mathbf k - {\bf K}_\perp)\,\sqrt{\frac{2\sinh(\pi\nu)}{(2\pi)^3\,\mathrm a\x\omega_{\,\bf K}}}\no
&\times\int_{0^+}^\infty\frac{\rmd x}{x}\;\e^{-\,iKx}\left(\x\nu + x\omega_{\,\bf K}\x\right)
K_{i\nu}(\kappa x)\no
& \equiv\;\alpha_{\x\mathbf K\x\mbox{\boldmath $\nu$}}
\end{align}
and in a quite similar manner
\begin{align}
(\,\bar u_{\,{\bf K}}^{\x\ast}\,,\,u_{\x\mbox{\boldmath $\nu$}}\,) &=
-\,\delta(\x{\bf k}+{\bf K}_\bot\x)\;\sqrt{\frac{2\sinh(\pi\nu)}{(2\pi)^3\,\mathrm a\x\omega_{\,\bf K}}}\no
&\times
\int_{0^+}^\infty\frac{\rmd x}{x}\,\e^{\,iKx}\,(\x\nu - x\omega_{\,\bf K}\x)\,\,K_{i\nu}(\kappa x)
\no
&\equiv\;\beta_{\x\mathbf K\x\mbox{\boldmath $\nu$}}
\end{align}
Now we turn to the calculation of the explicit expressions of the Bogoliubov
coefficients in terms of a suitable one parameter
representation.
From \cite{gradshteyn} 
we obtain
\begin{align}
\alpha_{\x\mathbf K\x\mbox{\boldmath $\nu$}}&=
-{\ts\frac{1}{2}}\left[\,\pi{\rm a}\,\omega_{\,\bf K}\,\sinh(\pi\nu)\,\right]^{\,-\,1/2}\,
\delta(\mathbf k - {\bf K}_\perp) \\
& \times \left(\frac{2\kappa}{\kappa+iK}\right)^{i\nu}\,\left\lbrace 
F\left({i\nu},i\nu+{\textstyle\frac12};{\textstyle\frac12};\frac{iK - \kappa}{\kappa+iK}\right)\right.\no
& \nn +\left. \frac{2\nu}{\omega_{\,\bf K}}(\kappa - iK)\,
F\left(1+{i\nu},i\nu+{\textstyle\frac12};{\textstyle\frac32};\frac{iK-\kappa}{iK+\kappa}\right)
\right\rbrace\\
\beta_{\x\mathbf K\x\mbox{\boldmath $\nu$}}&=
-\,{\ts\frac{1}{2}}\left[\,\pi{\rm a}\,\omega_{\,\bf K}\,\sinh(\pi\nu)\,\right]^{\,-\,1/2}\,
\delta({\bf k} + {\bf K}_\bot) \\
& \times \left(\frac{2\kappa}{\kappa-iK}\right)^{i\nu}\,\left\lbrace F\left({i\nu},i\nu+{\textstyle\frac12};{\textstyle\frac12};\frac{iK+\kappa}{iK-\kappa}\right)\right. \no
& \nn - \left. \frac{2\nu}{\omega_{\,\bf K}}(\kappa + iK)\,
F\left(1+{i\nu},i\nu+{\textstyle\frac12};{\textstyle\frac32};\frac{iK+\kappa}{iK-\kappa}\right)
\right\rbrace
\end{align}
where $F(a,b\x;c\x;z)$ denotes the hypergeometric function.
If we introduce the one angular parameter representation according to
\begin{align}
& \kappa+iK=\omega_{\,\bf K}\,\e^{\,i\theta_{\x\mathbf K}}
\qquad\qquad\theta_{\x\mathbf K}\equiv{\rm arctg}\frac{K}
{\surd\left( K_{\bot}^{2}+m^2\right) }
\no
&\frac{iK+\kappa}{iK-\kappa}\equiv\;(-1)\,\e^{\,2i\theta}
=\left(i\,\e^{\,i\theta}\right)^2
\qquad\qquad\ \left(\,-\,\textstyle\frac{\pi}{2}<\theta<\frac{\pi}{2}\,\right)
\end{align}
by making use of the the formul\ae
\begin{align}
& \Gamma(x)\Gamma\left(x+{\textstyle\frac12}\right)=\frac{2\sqrt{\pi}}{2^{\x2x}}\,\Gamma(2x) \\
& \nn (2k)!=\left({\textstyle\frac12}\right)_k\,k!\,2^{\x2k} \qquad (2k+1)!=\left({\textstyle\frac32}\right)_k\,k!\,2^{\x2k}
\end{align}
it is straightforward to get
\begin{align}
& F\left({i\nu},i\nu+{\textstyle\frac12};{\textstyle\frac12};-\,\e^{\x\pm\,2i\theta}\,\right)
=\sum_{k\,=\,0}^\infty\frac{\left(\,2i\nu\,\right)_{2k}}{(2k)!}\;\left(i\e^{\x\pm \,i\theta}\,\right)^{2k}\\
& -\;2\nu\mathrm e^{\x\pm \,i\theta}F\left(1+i\nu,i\nu+{\textstyle\frac12};{\textstyle\frac32};-\,\e^{\x\pm \,2i\theta}\,\right)= \\
&\nn \qquad\qquad\qquad\qquad\qquad =\sum_{k\,=\,0}^\infty\frac{\left(\,2i\nu\,\right)_{2k+1}}{(2k+1)!}\;\left(i\e^{\x\pm \,i\theta}\,\right)^{2k+1}
\label{hyper2}
\end{align}

Thus we are able to obtain exactly the sum of the above series in terms of elementary functions.
Consider in fact the identity
\begin{align}
F(a,b;b;z)&=\sum_{k=0}^\infty\frac{(a)_k}{k!}\,z^k=(1-z)^{-a}\no
&=\sum_{k=0}^\infty\frac{(a)_{2k}}{(2k)!}\,z^{2k}
+\sum_{k=0}^\infty\frac{(a)_{2k+1}}{(2k+1)!}\,z^{2k+1}
\end{align}
whence we can readily derive
\begin{align}
\sum_{k=0}^\infty\frac{(a)_{2k}}{(2k)!}\,z^{2k}&=
\textstyle\frac12\Big\{(1-z)^{-a}+(1+z)^{-a}\Big\}\\
\sum_{k=1}^\infty\frac{(a)_{2k-1}}{(2k-1)!}\,z^{2k-1}
&=\textstyle\frac12\Big\{(1-z)^{-a} - (1+z)^{-a}\Big\}
\end{align}
As a consequence, for $a=2i\nu$ and $z=(i\kappa \mp K)/\omega_{\,\bf K}$, 
we eventually succeed in extracting
the exact Bogoliubov coefficients for a spinless neutral massive field,
relating an inertial and an accelerated observers, viz.,
\begin{align}
& \alpha_{\x\mathbf K\x\mbox{\boldmath $\nu$}}\;=\no
&\frac{ -\,\delta(\x{\bf k} - {\bf K}_\perp)}
{\sqrt{4\pi{\rm a}\,\omega_{\,\bf K}\,\sinh(\pi\nu)}} \left\lbrace \frac{\omega_{\,\bf K}}{\omega_{\,\bf K} + K + i\kappa}\,\sqrt{
\frac{2\kappa}{\kappa + iK}}\;\right\rbrace ^{2i\nu} \no
&= -\,\delta(\x{\bf k} - {\bf K}_\perp)\;
\frac{\mathrm e^{\x\pi\nu/2}}{\sqrt{4\pi{\rm a}\,\omega_{\,\bf K}\,\sinh(\pi\nu)}}\;
\left( \frac{\omega_{\,\bf K} - K}{\omega_{\,\bf K} + K}\right) ^{i\nu/2}
\label{bogolalpha}\\
& \beta_{\x\mathbf K\x\mbox{\boldmath $\nu$}}\; =\no
&\frac{ -\,\delta(\x{\bf k} + {\bf K}_\perp)}
{\sqrt{4\pi{\rm a}\,\omega_{\,\bf K}\,\sinh(\pi\nu)}}
\left\lbrace \frac{\omega_{\,\bf K}}{\omega_{\,\bf K} + K - i\kappa}\,\sqrt{
\frac{2\kappa}{\kappa - iK}}\;\right\rbrace ^{2i\nu}\no
&= -\,\delta(\x{\bf k} + {\bf K}_\perp)\;
\frac{\mathrm e^{\x-\,\pi\nu/2}}{\sqrt{4\pi{\rm a}\,\omega_{\,\bf K}\,\sinh(\pi\nu)}}\;
\left( \frac{\omega_{\,\bf K} - K}{\omega_{\,\bf K} + K}\right) ^{i\nu/2}
\label{bogolbeta}
\end{align}
which are in full agreement, up to the Dirac $\delta-$distribution in the tranverse wave vectors,
with eq.s~(2.107) and (2.108) of the review paper \cite{crispino}.
It follows that, keeping apart the classical volume factors
\begin{equation}
\Omega_{\x\mathbf K}^{\x(\pm)}=
\delta(\x{\bf k} \pm {\bf K}_\perp\x)\;\sqrt\frac{c^{\x3}}{2\pi{\rm a}\,\omega_{\,\bf K}}
\end{equation}
we can understand the square modulus of the complex dimensionless quantities
\begin{equation}
 \frac{-\;\mathrm e^{\x\pm\,\pi\nu/2}}{\sqrt{2\sinh(\pi\nu)}}\;
\left( \frac{\omega_{\,\bf K} + K}{\omega_{\,\bf K} - K}\right) ^{-\,i\nu/2}
\end{equation} 
as nonnegative average numbers. As a matter of fact we set
\begin{eqnarray}
 \mathrm N_{\x\nu}\equiv
\frac{e^{\x2\pi\nu}}{\mathrm e^{\x2\pi\nu} - 1}
\qquad\quad
 \bar{\mathrm N}_{\x\nu}\equiv
\frac{1}{\mathrm e^{\x2\pi\nu} - 1}
\end{eqnarray}
and if we definitely understand as it is customary
$2\pi\nu=\beta E$,
where $\beta=(2\pi c/\hbar\rm a)$ so that
$T=(\hbar{\rm a}/2\pi ck_{\x\rm B})$ is the famous Unruh temperature \cite{Unruh},
$k_{\x\rm B}$ being the Boltzmann constant, we eventually recover the thermal
bath which is experienced by the accelerated observer in respect to the
inertial one or \textit{vice versa.} Moreover, from the knowledge of the Bogoliubov coefficients
one can immediately obtain the transformation law for the creation and destruction operators
\begin{align}
& \bar a_{\x\mathbf K}=\sum_{\x\mbox{\boldmath $\nu$}}
\left[\, a_{\x\mbox{\boldmath $\nu$}}\,\alpha_{\x\mathbf K\x\mbox{\boldmath $\nu$}}
- \beta_{\x\mathbf K\x\mbox{\boldmath $\nu$}}^{\x\ast}	\, a^{\x\dagger}_{\x\mbox{\boldmath $\nu$}}\,\right] \\
\nn & \bar a^{\x\dagger}_{\x\mathbf K}=\sum_{\x\mbox{\boldmath $\nu$}}
\left[\, a^{\x\dagger}_{\x\mbox{\boldmath $\nu$}}\,\alpha^{\x\ast}_{\x\mathbf K\x\mbox{\boldmath $\nu$}}
- \beta_{\x\mathbf K\x\mbox{\boldmath $\nu$}}\,a_{\x\mbox{\boldmath $\nu$}}\,\right]
\end{align}
where
\begin{equation}
 \sum_{\x\mbox{\boldmath $\nu$}}\equiv\mathrm a\int_{-\infty}^{\infty}\mathrm d\nu\int\mathrm d^2k
\end{equation}
The above derivation, that concerns the spinless neutral
field in the right Rindler wedge, can be trivially extended to the left Rindler wedge.
Furthermore, our direct method of evaluation for the  Bogoliubov coefficients
is well tailored to be successfully generalized \textit{mutatis mutandis} to fields of higher spin
and any charge. Finally,
concerning the recent debate \cite{fedotov,FU} on the validity of the Unruh effect,
our first principles derivation of the Bogoliubov coefficients definitely endorses the claim of \cite{FU}, 
in a manner that is substantially different from Unruh's original trick \cite{Unruh}.

\begin{acknowledgments}
R.S. wishes to acknowledge the support of the Istituto Nazionale di Fisica Nucleare,
IS PI13, that contributed to the successful completion of this project.
\end{acknowledgments}

%
%

\providecommand{\noopsort}[1]{}\providecommand{\singleletter}[1]{#1}%

\end{document}